\documentclass[11pt]{article}

\usepackage{mathptmx}
\usepackage{amsmath,amsfonts,amsthm,amssymb,latexsym,color,palatino}
\usepackage[colorlinks=true,urlcolor=webblue,linkcolor=webgreen,filecolor=webblue,citecolor=webgreen,pdfpagemode=UseOutlines,pdfstartview=FitH,pdfpagelayout=OneColumn,bookmarks=true]{hyperref}
\usepackage{fullpage}

\hypersetup{
  pdftitle=Quantum Algorithms for Simon's Problem Over General Groups,
  pdfauthor=Gorjan Alagic and Cristopher Moore and Alexander Russell}

\definecolor{webgreen}{rgb}{0,.5,0}
\definecolor{webblue}{rgb}{0,0,.5}

\newtheorem{prop}{Proposition}

\newcommand\identity{\leavevmode\hbox{\small1\kern-3.2pt\normalsize1}}

\newcommand\ZZ{\mathbb Z}
\newcommand\CC{\mathbb C}
\newcommand\EE{\mathbb E}
\newcommand\U{\textsf{U}}

\newcommand{\kett}[1]{\left| #1 \right\rangle}
\newcommand{\bbra}[1]{\left\langle #1 \right|}

\newcommand{\remove}[1]{}
\newcommand{\e}{{\mathrm e}}

\newcommand{\wk}{\widehat{K}}
\newcommand{\tr}{{\rm tr}\ }
\newcommand{\sign}{{\chi_\pm}}

\newcommand{\id}{\bar{1}}

\begin{document}
\title{Quantum Algorithms for Simon's Problem\\ Over General Groups} 

\author{Gorjan Alagic\\
  University of Connecticut\\
  \texttt{alagic@math.uconn.edu} \and
  Cristopher Moore\thanks{Supported by NSF grants CCR-0220070, EIA-0218563, and CCF-0524613, and ARO-ARDA grant 47976-PH-QC.}\\
  University of New Mexico\\
  \texttt{moore@cs.unm.edu}
  \and
  Alexander Russell\thanks{Supported by NSF CAREER award CCR-0093065, NSF grants EIA-0523456 and EIA-0523431, and ARO-ARDA grant 47976-PH-QC.}\\
  University of Connecticut\\
  \texttt{acr@cse.uconn.edu}}

\maketitle
\begin{abstract}
  Daniel Simon's 1994 discovery of an efficient quantum algorithm for
  solving the hidden subgroup problem (HSP) over $\ZZ_2^n$ provided
  one of the first algebraic problems for which quantum computers are exponentially faster
  than their classical counterparts. In this paper, we study the generalization of Simon's problem to
  arbitrary groups.  Fixing a finite group $G$, this is the problem of
  recovering an involution $\bar{m} = (m_1, \ldots, m_n) \in G^n$ from an
  oracle $f$ with the property that $f(\bar{x})   f(\bar{x} \cdot \bar{y}) \Leftrightarrow \bar{y} \in \{ \id, \bar{m} \}$.  
  In the current parlance, this is the hidden subgroup problem (HSP) 
  over groups of the form $G^n$, where $G$ is a nonabelian group of constant size, 
  and where the hidden subgroup is either trivial or has order two.  
  
  Although groups of the form $G^n$ have a simple product structure, they 
  share important representation-theoretic properties with the symmetric groups $S_n$, 
  where a solution to the HSP would yield a quantum algorithm for Graph Isomorphism. 
  In particular, solving their HSP with the so-called ``standard method''  requires 
  highly entangled measurements on the tensor product of many coset states.

  Here we give quantum algorithms with time complexity $2^{O(\sqrt{n \log n})}$ that 
  recover hidden involutions $\overline{m} = (m_1, \ldots, m_n) \in G^n$ where, 
  as in Simon's problem, each $m_i$ is either the identity or the conjugate of a
  known element $m$ and there is a character $\chi$ of $G$ for which $\chi(m) = -\chi(1)$.
  Our approach combines the general idea behind Kuperberg's sieve for
  dihedral groups with the ``missing harmonic'' approach of Moore and
  Russell.  These are the first nontrivial hidden subgroup algorithms
  for group families that require highly entangled multiregister
  Fourier sampling.
\end{abstract}

\section{Introduction}

\subsection{Simon's problem over general groups}

Let $G$ be a finite group of even order and suppose we are given access to an oracle 
function $f$, from $G^n$ to some set $S$, with the property that 
$f(\bar x \cdot \bar y) = f(\bar x) \Leftrightarrow \bar{y} \in \{ \id, \bar{m} \}$ 
for some ``hidden'' involution $\bar{m} \in G^n$. Our task is to determine $\bar{m}$ by querying the oracle $f$. 
While this problem is classically intractable, a beautiful quantum algorithm of Daniel Simon's
solves it efficiently when $G \cong \ZZ_2$~\cite{Simon}. 
In this article, we study this problem over general groups. This problem fits into the
framework of the \emph{Hidden Subgroup Problem} (HSP), which also underlies 
Shor's celebrated quantum algorithms for factoring and discrete logarithm~\cite{Shor}.  
In general, an instance of a hidden subgroup problem over a group $K$ is described by
a function $f: K \to S$ that ``hides'' a subgroup $H \leq K$ in the sense that
\begin{equation}\label{eq:oracle}
f(k) = f(k \cdot h) \Leftrightarrow h \in H \enspace .
\end{equation}
The problem is to determine the subgroup $H$ by making queries to the
function $f$. For the cases of interest in this article, the group $K
= G^n$ is exponentially large, and we measure the complexity of
algorithms for the problem as a function of $\log |K| = \Theta(n)$.
Moreover, the subgroup $H$ is of constant size, making it easy to rule
out the possibility of efficient classical algorithms.

The number-theoretic problems solved by Shor's algorithms are naturally related to
the HSP on \emph{abelian} groups which are, at this point, quite well
understood.  The HSP on \emph{nonabelian} groups has very exciting
algorithmic applications, including Graph Isomorphism and certain
lattice problems~\cite{Regev}; however, these problem appear to
be quite challenging in general.  Efficient algorithms exist for a number of families
of nonabelian groups
(e.g.,~\cite{RoettelerB98,IvanyosMS01,FriedlIMSS02,MooreRRS04,InuiLeGall,BaconCvD2})
but all of these are ``nearly abelian'' in one sense or another.
In contrast, the groups of greatest interest, the symmetric groups
$S_n$ whose hidden subgroup problems correspond to Graph Isomorphism,
have steadfastly resisted the community's advances.  A series of negative
results~\cite{HallgrenRT00,MRS05,HallgrenMRRS} have shown that the
standard approach of measuring a \emph{coset state}, i.\,e., a uniform
superposition over a random coset of $H$, is inherently limited.  The
strongest of these results, due to Hallgren et al.~\cite{HallgrenMRRS} 
shows that even to obtain enough \emph{information} to solve the HSP (regardless of 
the computational complexity of processing this information) any
algorithm based on this approach must involve highly entangled measurements
on the tensor product of $\Omega(\log |S_n|)=\Omega(n \log n)$ coset states.

In spite of the fact that groups of the form $G^n$ have a simple direct product structure, 
all the known negative results that apply to $S_n$ apply in this setting as well.  
In particular, whenever $G$ is nonabelian almost all the representations of $G^n$ 
have dimensions which are exponentially large, causing strong Fourier sampling 
of single coset states to fail~\cite{AMR05}.  Moreover, solving Simon's problem on $G^n$ 
requires highly entangled measurements over $\Omega(\log |G^n|) = \Omega(n)$ coset states.  

In this article, we describe a family of algorithms for Simon's
problem on $G^n$ so long as there is a character $\chi$ of $G$ which, as in 
the original problem on $\ZZ_2^n$, ``identifies'' nontrivial coordinates of the involution. Specifically, we recover hidden subgroups $H$ of the form $\{ 1, \bar{m}\}$ where, as in Simon's original problem, (i.) each $m_i$ is either the identity or the conjugate of a known element $\mu \in G$ and (ii.) there is a character $\chi$ of $G$ for which $\chi(\mu) = -\chi(1)$. Our algorithms take
time $2^{O(\sqrt{n \log n})}$.  We emphasize that these are the first
subexponential-time algorithms to which the negative results
of~\cite{MRS05,HallgrenMRRS} apply, and for which highly entangled
measurements are necessary.  

Our algorithms use an adaptive sieve similar 
to that employed by Kuperberg in his algorithm for the dihedral 
groups~\cite{Kuperberg}. Specifically, we combine registers pairwise in 
an effort to generate states lying in one-dimensional representations.  
The rules by which we match pairs of registers together are more complicated 
than those in the dihedral group, however, and depend on the Clebsch-Gordan problem in $G$.
Ultimately, our algorithms produce states which lie in representations which are chosen 
uniformly from the one-dimensional representations of $G^n$ which are 
orthogonal to the hidden subgroup.  Their success rests on the fact that, given any
order-2 subgroup $H$, at least half of these representations are ``missing harmonics'' 
of $H$ and cannot appear~\cite{Banff} unless the hidden subgroup is in fact trivial.

\subsection{The representations of $G^n$}
\label{sec:repsofgn}

Recall that a \emph{representation} of a group $K$ is a homomorphism
$\rho:K \to \U(V_\rho)$, where $\U(V_\rho)$ is the set of unitary operators acting on
some vector space $V_\rho$, whose dimension we denote by $d_\rho$.  A representation is
\emph{irreducible} if there are no proper subspaces of $V_\rho$ which are
preserved by $\rho(g)$ for all $g \in K$.  
Schur's Lemma asserts that for any irreducible representation (or ``irrep'') $\rho$, 
the only operators which commute with its image $\text{Im}\ \rho \in \U(V)$ are the scalar matrices $\lambda \identity$ where $\lambda \in \CC$.

We say that two irreps are \emph{isomorphic} if there
is a unitary change of basis that maps one onto the other; that is,
$\rho_1 \cong \rho_2$ if there is a $T$ such that $\rho_1(g) = T^\dagger \rho_2(g) T$ for
all $g \in K$.  If $K$ is finite, there are a finite number of
equivalence classes under isomorphism, and we let $\wk$ denote 
one irrep from each of these equivalence classes.  Given an irrep $\rho$,
its \emph{character} at a group element $k$ is defined to be $\chi_\rho(k) = \tr \rho(k)$, 
and one-dimensional irreps can be identified with their characters.

Once a basis is chosen for the spaces $V_\rho$ of each $\rho \in \wk$, 
the $\rho(g)$ become explicit matrices, and the matrix elements of the irreps in $\wk$ form
a basis for $\CC K$, the algebra of complex-valued functions on $K$.  The
\emph{quantum Fourier transform} is the change of basis from the group
basis 
\[
\{ \kett{k} : k \in K \}
\]
of $\CC G$ to the Fourier basis 
\begin{equation}\label{eq:Fourier-basis}
\{ \kett{\rho,i,j} : \rho \in \wk \text{ and } 1 \leq i,j \leq d_\rho\}\enspace.
\end{equation}

If $\rho: K \to \U(V)$ is a \emph{reducible} representation, one may find an
orthogonal decomposition $V_\rho = V_1 \oplus \cdots \oplus V_k$ with the remarkable
property that every $\rho(g)$ preserves each subspace $V_i$ and,
moreover, that the matrices $\rho_i(g)$ obtained by restricting $\rho(g)$ to
$V_i$ form an irreducible representation.  In this case we write $\rho \cong
\oplus_i \rho_i$. Note that, up to isomorphism, a given representation $\tau \in \wk$
may appear many times in such a decomposition. 

We remark that if $\rho$ and $\sigma$ 
are representations of a group $K$, one may naturally define a representation $\rho \otimes \sigma$ 
on the space $V_\rho \otimes V_\sigma$ by the ``diagonal action'': the matrix $[\rho \otimes \sigma](g)$ 
is defined to be $\rho(g) \otimes \sigma(g)$. Even when $\rho$ and $\sigma$ are irreducible, 
the representation $\rho \otimes \sigma$ is, in general, reducible; the problem of determining the
irreducible constituents of $\rho \otimes \sigma$ is the \emph{Clebsch-Gordan}
problem.

Finally, we mention that the representation theory of a product group decomposes into that of
its factors. In particular, each irrep $\bar \rho$ of $G^n$ is a tensor
product $\rho_1 \otimes \cdots \otimes \rho_n$ of $n$ irreps of $G$. Specifically,
$\bar \rho(\bar x) = \rho_1(x_1) \otimes \cdots \otimes \rho_n(x_n)$ (note that this differs from
the ``diagonal'' tensor product construction described above, since the components of $\bar x$ 
may differ). 
The dimension $d_{\bar \rho}$ of $\bar \rho$ is hence equal to the product of the
dimensions of the $\rho_i$, and the character $\chi_{\bar \rho}(\bar x)$ is
equal to the product $\chi_{\rho_1}(x_1) \cdots \chi_{\rho_n}(x_n)$ of the
characters of the $\rho_i$ at the $x_i$.

\section{Previous negative results}

\subsection{Weak sampling fails}

The standard approach to the HSP is to create a uniform quantum superposition over the group, 
and then measure the value of the oracle $f$.  This produces the so-called \emph{coset state}, 
which is a uniform superposition over a random coset of the hidden subgroup $H$.  

Using the techniques of~\cite{MRR03}, it is easy to perform the quantum Fourier transform 
for groups of the form $G^n$ in ${\rm poly}(n)$ time.  
The process of \emph{weak Fourier sampling} then consists of measuring just the ``name'' 
of the representation ${\bar \rho} \in \widehat{G^n}$ (that is, the leading portion of the register in the Fourier basis~\eqref{eq:Fourier-basis}), while \emph{strong Fourier sampling} 
also measures the row and column $i,j$ in a basis of our choice.  
We begin by discussing the probability distribution $P_H(\bar \rho)$ on $\widehat{G^n}$ 
observed under weak sampling.   This is given by
\begin{equation}
\label{eq:weak}
  P_H(\bar \rho) = \frac{d_{\bar \rho}|H|}{|G|^n} {\bf rk}~\Pi_H^{\bar \rho}\enspace,
\end{equation}
where 
\[ \Pi_H^{\bar \rho} = \frac{1}{|H|} \sum_{h \in H} {\bar \rho}(h) \] 
is the projection operator formed by averaging the
representation $\bar \rho$ over $H$.  When $H = \{\id\}$, this is
$$
 P_{\{\id\}}(\bar \rho) = \frac{d_{\bar \rho}^2}{|G|^n}\enspace,
$$
often referred to as the \emph{Plancherel} distribution; this is simply 
the dimensionwise fraction of $\CC G^n$ consisting of copies of 
$\bar \rho$. The following proposition, proved in \cite{AMR05}, asserts that if 
the ``base group'' $G$ possesses an involution outside its center $Z(G)$, then 
weak sampling cannot solve Simon's problem on $G^n$: that is, there are 
many involutions ${\bar{m}}$ which it cannot distinguish from each other, and indeed 
it cannot distinguish the order-2 subgroup $\{\id,\bar{m}\}$ from the trivial subgroup.

\begin{prop} \label{weak-samp-thm} Let $G$ be a group with an
  involution $\mu \notin Z(G)$, and let $H = \{\id, (m_1, \ldots, m_n)\} \leq G^n$, 
  where each $m_i$ is conjugate to $\mu$. Then the total variation distance 
  between the weak Fourier sampling distributions for $H$ and $\{\id\}$ is at most $2^{-n/2}$.
\end{prop}

\subsection{Strong sampling fails}

Using the machinery of Moore, Russell, and Schulman~\cite{MRS05}, one
can show that performing strong Fourier sampling on a single coset state  
also fails to solve Simon's problem on $G^n$ for many choices of $G$.   
Given a suitable condition on $G$, a simple application of a Chernoff bound shows that 
the dimension of the irrep $\bar \rho$ resulting from weak Fourier sampling is 
exponentially large in $n$ with overwhelming probability.  It is not
hard to use this fact and the results of~\cite{MRS05} 
to show that single-register strong Fourier sampling then fails
for $G^n$; for details, see the present authors' preprint~\cite{AMR05}.  

For a slightly stronger condition on $G$, Hallgren, Moore, R\"otteler, Russell and Sen~\cite{HallgrenMRRS} 
recently showed that entangled measurements over $\Omega(n)$ registers are required to distinguish subgroups of $G^n$.  
This result applies to many interesting base groups, including all nonabelian simple groups and the 
symmetric groups $S_k$ for $k \geq 4$.

These negative results strongly suggest that there is little hope of developing a generic recursive 
approach to the HSP based on breaking the group into a tower of subgroups.  Not only can we decompose 
$G^n$ into the tower $G^n > G^{n-1} > \cdots > \{1\}$, but if $G$ is \emph{solvable} we can further 
refine this tower so that each subgroup is normal and all the quotient groups are abelian.  
For certain solvable groups, Friedl et al.~\cite{FriedlIMSS02} showed that such a tower allows us 
to solve the HSP recursively.  However, there are groups to which the results of~\cite{AMR05,HallgrenMRRS} 
apply which are solvable and even \emph{nilpotent}, such as the ``chandelier groups'' formed by the 
$k$-fold wreath product $\ZZ_2 \wr \cdots \wr \ZZ_2$ and which describe the automorphisms of a binary 
tree of depth $k$.

\section{An algorithm for the product groups $S_k^n$}

\subsection{Outline of the algorithm}

In this section, we outline our algorithm for Simon's problem on $G^n$ for $G$ of constant size.  
For concreteness, we will describe the case where $G$ is the symmetric group $S_k$. Our algorithm will 
determine the hidden subgroup $H \leq S_k^n$, given oracle access to a function $f$ 
satisfying~\eqref{eq:oracle}, and the promise that $H$ is either the trivial subgroup $\{\id\}$ or of the form 
$\{\id, \bar{m}\}$ where $\bar{m} = (m_1, m_2, \dots, m_n)$ with each $m_i$ an odd involution in $S_k$. Note that for any odd permutation $\pi \in S_k$, $\sign(\pi) = -1 = -\sign(1)$, where $\sign: S_k \rightarrow \CC$ is the \emph{sign} representation mapping each permutation to its sign. We explain how to generalize the algorithm to other base groups and subgroups in 
Section~\ref{sec:generalize}.

For enhanced readability, we make two additional simplifications in this section.
First, our algorithm will only determine if $H$ is trivial or nontrivial; in
Section~\ref{sec:determining} we show how to determine the involution $\bar{m}$. 
Second, we put off detailed descriptions of the quantum computational 
ingredients of the algorithm until Section~\ref{sec:weaksamp}, choosing instead to simply list them here
and outline the algorithm in terms which require only a basic knowledge of
representation theory. These ``quantum ingredients'' are efficient in
the sense that they require time at most polynomial in $n$. We thus
assume that:
\begin{enumerate}
\item We can use the oracle to sample irreps $\bar \rho$ of $S_k^n$ according to the probability
distribution~\eqref{eq:weak}.  In particular, if $H$ is the trivial subgroup, it will sample irreps according to the Plancherel distribution. 
\item Given two irreps $\bar \rho$ and $\bar \sigma$ of $S_k^n$, we can
  ``Fourier sample'' inside $\bar \rho \otimes \bar \sigma$. When $H$ is the trivial
  subgroup this results in an irrep $\bar \tau$ according to the
  natural distribution induced by its appearance in the 
  decomposition of $\bar \rho \otimes \bar \sigma$ into $G^n$-irreps.  
  When $H$ is nontrivial, the distribution depends both on $H$ and the history of the algorithm up to that point.
\item Finally, if we can use (1) and (2) to produce a \emph{missing harmonic}~\cite{Banff}, i.e, one for which $\bar \rho (\bar{m}) = -\identity$, 
 then $H$ is the trivial subgroup.  These are exactly the one-dimensional 
 $\bar \rho$ such that the number of $\rho_i$ 
 isomorphic to the sign representation of $S_k$ is odd.  On the other hand, producing a one-dimensional 
 $\bar \rho$ with an even number of sign representations provides evidence that $H$ is nontrivial.
\end{enumerate}

\noindent 
Our algorithm will first use (1) to produce a large number of irreps of $S_k^n$, 
then use (2) repeatedly in a ``sieve'' to whittle our irreps down to one-dimensional ones, and finally apply (3) to make a correct guess about $H$ with overwhelming probability.

We now discuss our method for selecting irreps to combine via step (2).
Suppose we have the tensor product of two irreps $\bar \rho$ and $\bar \sigma$ of $S_k^n$.  
We wish to decompose them into a direct sum of irreps of $S_k^n$, 
$$
\bar \rho \otimes \bar \sigma 
%= \bigotimes_{i=1}^n (\rho_i \otimes \sigma_i) 
= \bigoplus {\bar \tau} \enspace . 
$$ 
These $\bar \tau$ are simply tensor products of the $\tau_i$ appearing in the decomposition 
of $\rho_i \otimes \sigma_i$ into irreps of $S_k$ for each $i$.  Now, if $d_{\rho_i} = d_{\sigma_i} = 1$, 
then clearly $\rho_i \otimes \sigma_i$ consists of a single one-dimensional irrep.  It is a fact of 
the representation theory of the symmetric group that if $\sigma_i$ is isomorphic to $\rho_i$, then $\rho_i \otimes
\sigma_i$ contains one copy of the trivial representation; likewise, if $\sigma_i$ is isomorphic to 
the conjugate representation $\rho_i^\perp$ (obtained by flipping $\rho_i$'s Young diagram across its 
main diagonal) then $\rho_i \otimes \sigma_i$ contains one copy of the sign representation.  

Hence, we would ideally like to use (2) to combine irreps $\bar \rho$ and $\bar \sigma$ which match up 
in a nice way, i.\,e., such that $\rho_i$ is one-dimensional exactly when $\sigma_i$ is one-dimensional, 
and such that either $\sigma_i \cong \rho_i$ or $\sigma_i \cong \rho_i^\perp$ for all other indices $i$.  
In that case, if $H$ is trivial, each $\tau_i$ would be one-dimensional with constant probability.  
However, producing such pairs of irreps of $S_k^n$ simply using (1), i.\,e., by independently weak sampling 
each one, would require exponential time.  Instead, we work our way towards missing harmonics with a 
sieve in the spirit of Kuperberg's algorithm for the dihedral groups~\cite{Kuperberg}.

Specifically, we will use (1) to produce a pool of $2^{\Theta(\sqrt{n \log n})}$
$S_k^n$-irreps, and then repeatedly pairwise combine them using (2)
according to the following rule.  First, for each $\bar \rho$ in our
pool, we flip $\sqrt{n / \log n}$ coins, setting $z^{\bar \rho}_j=0$ or $1$ with
equal probability for $j = 1, \ldots, \sqrt{n / \log n}$.  We then consider
the $\sqrt{n / \log n}$ leftmost indices $i$ for which $d_{\rho_i} > 1$.
We say that $\bar \sigma$ is a \emph{partner} for $\bar \rho$ if, for the $j$th such
index $i$, we have $\rho_i \cong \sigma_i$ or $\rho_i \cong \sigma_i^\perp$ if $z^{\bar \rho}_j=0$ or $1$
respectively, and moreover if $z^{\bar \rho}_j = z^{\bar \sigma}_j$ for all $j$. 
Observe that there are then at most
\begin{multline*}
\binom{n}{\sqrt{n / \log n}} (2 |\widehat {S_k}|)^{\sqrt{n / \log n}} \\
\approx \left(\frac{\e n}{\sqrt{n / \log n}}\right)^{\sqrt {n / \log n}} (2 |\widehat {S_k}|)^{\sqrt{n / \log n}} \\
= 2^{\Theta(\sqrt {n \log n})}
\end{multline*}
different types of $S_k^n$-irreps for the purposes of this method of
matching.  Our algorithm relies on the fact that
if $\bar \tau$ is produced from $\bar \rho \otimes \bar \sigma$ using (2), then
whenever $\rho_i \cong \sigma_i$ or $\rho_i \cong \sigma_i^\perp$, we will have $d_{\tau_i} = 1$ with
constant probability.

The following is an outline of our algorithm.

\begin{itemize}

\item Use quantum ingredient (1) to produce a collection $\Lambda$ of $2^{\Theta(\sqrt{n \log n})}$ irreps of $S_k^n$. 

\item Repeat $6 \sqrt{n \log n}$ times:

\begin{itemize}

\item Pair up the irreps from $\Lambda$ according to the rule specified above, and discard any
      unpaired irreps.

\item Use quantum ingredient (2) to combine each pair of partners to produce a new $S_k^n$-irrep. Redefine $\Lambda$ to be the set of ``children'' produced in this manner.

\end{itemize}

\item By quantum ingredient (3), we may output ``trivial'' if $\Lambda$ contains any missing harmonics, and ``nontrivial'' otherwise.  

\end{itemize}

\noindent Since we never discard more than a constant fraction of the irreps at each step, 
it is clear that we will still have the necessary $2^{\Theta(\sqrt {n \log n})}$ irreps at the last step. 
If $H$ is trivial, it is easy to see that every one-dimensional irrep 
has the same probability of appearing in the final list $\Lambda$ since our coin-flipping process 
chooses partners where $\sigma_i \cong \rho_i$ or $\rho_i^\perp$ with equal probability.  
Since half of the possible one-dimensional irreps are missing harmonics, the probability that 
$\Lambda$ contains no missing harmonics in this case is superexponentially small.  
Hence the probability the final collection contains a missing harmonic is $1-o(1)$ if $H$ is trivial,
and $0$ if it is not. The output is thus correct with high probability.

\subsection{Analysis}

Suppose $\bar \tau$ is a child resulting from applying ingredient (2) to a pair of partners.  
Let the \emph{weight} of $\bar \tau$ be the number of its factors $\tau_i$ with $d_{\tau_i} > 1$.  
Ideally, we would like $\bar \tau$ to satisfy one of the following three conditions: 1) the weight 
of $\bar \tau$ is zero; 2) the parents have weight at least $\sqrt{n / \log n}$,
and the weight of $\bar \tau$ is at least $c_1 \sqrt{n / \log n}$ lower; 3)
the parents have weight less than $\sqrt{n / \log n}$ and the weight of $\bar \tau$ is lower by a 
constant fraction $c_2$. Since we know the dimensions of the irreps 
of $S_k$ (which is of constant size), it is easy to select the constants $c_1$ and $c_2$ such that one of these three
conditions is satisfied with probability at least $1/2$, for every child produced
in this manner. In the following, we assume that these constants have been 
chosen as prescribed, and that the algorithm is run for $m = 6 \sqrt{n \log n}$ steps.

Let $\bar \rho_0$ be an irrep of $S_k^n$ sampled at the beginning of the algorithm, and 
let $\{\bar \rho_j\}_{j=1}^m$ be the sequence consisting of the descendants of $\bar \rho_0$ given in
the obvious order (e.g.~$\bar \rho_2$ is the child of $\bar \rho_1$, which is the child of $\bar \rho_0$.)
Note that we are assuming that we have selected an irrep produced
initially via ingredient (1) which has a surviving descendant in the final collection $\Lambda$.
Our goal is to prove that $\bar \rho_m$ is one-dimensional (i.\,e., has weight zero) with
very high probability. Define a sequence of random variables $\{A_j\}_{j=1}^m$ such that $A_j = 1$ if
$\bar \rho_j$ satisfies one of the three conditions outlined above, and $A_j = 0$ otherwise.
Notice that
$$
\EE \Bigl[ \sum_{j=1}^m A_j \Bigr] \geq \frac{m}{2} = 3\sqrt{n \log n}\enspace,
$$
but that $\bar \rho_m$ has weight zero unless 
$$
\sum_{j=1}^m A_j < \sqrt{n \log n} + O(\log n)\enspace.
$$
We now define
the random variables
$$
C_k = \EE \Bigl[  \Bigl. \sum_{j=1}^m A_j ~\Bigr|~ \{ A_i : i \leq k \} \Bigr] \qquad k = 0, 1, 2, \dots, m\enspace. 
$$
We will control the quantity
$$
\bigl| C_m - C_0\bigr| = \left|\sum_{j=1}^m A_j - \EE \Bigl[ \sum_{j=1}^m A_j \Bigr] \right|\enspace.
$$
Observe that $\{C_k\}_{k=0}^m$ defines a (sub)martingale, 
i.\,e., $\EE \left[ C_{k+1} \mid C_k \right] \geq C_k$, and that $|C_{k+1} - C_k| \leq 1$.
We can thus apply Azuma's inequality, which asserts that
$$
\text{Pr}\left[\left|C_m - C_0\right| \geq \sqrt{n \log n} \right] \leq 2 \e^{-\sqrt{n \log n}/12}\enspace,
$$
and conclude that, with overwhelming probability,
$$
\sum_{j=1}^m A_j \geq 2 \sqrt{n \log n}
$$
and the weight of $\bar \rho_m$ is zero, as desired.

\subsection{Weak sampling tensor product states and missing harmonics}\label{sec:weaksamp}

Quantum ingredient (1) of our algorithm is the familiar tool of weak Fourier sampling.  
We continue here with a discussion of the slightly less familiar quantum ingredients (2) and (3), 
for a general group $G$ and subgroup $H$.  Recall that at the beginning of the algorithm, each register contains the mixed coset state
$$
\rho_H = \frac{1}{|G|} \sum_{c \in G} |cH \rangle \langle cH|\enspace,
$$
where $\kett{S}$ denotes the state 
$$
\frac{1}{\sqrt{|S|}} \sum_{s \in S} \kett{s}\enspace;
$$
note that this is the completely mixed state in the case where $H$ is trivial.  
We then apply weak Fourier sampling to each register, projecting it
into a right-invariant irreducible subspace of $\CC G$ corresponding
to some irrep $\rho$ of $G$. We remark at this point that the density matrix 
$\rho_H$ is invariant under right multiplication by any element of $H$;
in particular,
\begin{equation}
\label{eq:invariance}
R_h \rho_H = \rho_H R_h = \rho_H
\end{equation}
where $R_h$ is the unitary operator $R_h: \kett{g} \mapsto \kett{gh}$ 
corresponding to (right) multiplication by an element $h \in H$.  
Note that this is stronger than the property that $\rho_H$ commutes with $R_h$, which 
is another sense in which a mixed state can be invariant under a unitary operator.

The ``combine'' step of our algorithm consists of the following sampling procedure, 
identical to that used by Kuperberg~\cite{Kuperberg} for the dihedral groups.  First,
a pair of these weakly sampled registers can be described as a state in the vector space 
$$
V_\rho \otimes V_\sigma \subset \CC G \otimes \CC G
$$
where $V_\rho$ and $V_\sigma$ are irreducible right-invariant spaces of $\CC G$ 
corresponding to irreps $\rho$ and $\sigma$.  
When $V_\rho \otimes V_\sigma$ is treated as a representation of $G$ under the
right diagonal action it is typically reducible, as stated in Section~\ref{sec:repsofgn}: 
its Clebsch-Gordan decomposition
is the direct sum $V_\rho \otimes V_\sigma \cong \oplus_\tau W_\tau$, where, for an irrep
$\tau$, $W_\tau$ is the span of all subspaces of $V_\rho \otimes V_\sigma$ isomorphic to
$\tau$.  We call such a subspace \emph{isotypic}.  
Our goal is to perform ``isotypic sampling'' on this tensor product state: in other words, to 
measure the irrep name $\tau$ corresponding to the isotypic subspace $W_\tau$.

If $\kett{\psi}$ is our original state in $V_\rho \otimes V_\sigma$, let $\kett{\psi_\tau}$ denote 
the projection of $\kett{\psi}$ into $W_\tau$.
We begin by applying the controlled-multiplication operation to the
entire space $\CC G \otimes \CC G$ defined by
$$
M : |a \rangle | b \rangle \mapsto |a \rangle | ba^{-1} \rangle.
$$
As pointed out in~\cite{Kuperberg}, this unitary transformation transports 
the right diagonal $G$-action to the $G$-action on the first register.  
Therefore, we can now apply the quantum Fourier transform to the first
register of the result and measure a representation name $\tau$.  The
portions of this first register corresponding to a state in a right
$G$-invariant space (a ``column'' of $\tau$) are kept; all other portions
of this register and the right hand register are left unmeasured (and
unused!). By computing the internal trace across these qubits, we see
that the resulting mixed state is precisely the one we wanted: the
irrep name $\tau$ is measured with probability $\| \kett{\psi_\tau} \|^2$.
Moreover, as promised in ingredient
(2), when $H$ is the trivial subgroup we have the fully mixed state at
each point of the algorithm and hence the probability of observing $\tau$
is precisely $\dim W_\tau / (\dim V_\rho \cdot \dim V_\sigma)$, the dimensionwise
fraction of $V_\rho \otimes V_\sigma$ composed of copies of $\tau$. This establishes
the claims of ingredient (2).

Note, however, that when the subgroup $H$ is nontrivial the
distribution induced by the observation above may be rather
complicated, depending both on the subgroup $H$ and the exact sequence
of measurements that resulted in the states present in $V_\sigma$ and
$V_\tau$. The only fact we shall use about the states produced at
intermediate stages of the algorithm for nontrivial $H$, however, is
that the combine step preserves invariance under right multiplication
by elements of $H$.  Indeed, if $\alpha_\rho$ and $\alpha_\sigma$ are mixed states over
the spaces $V_\rho$ and $V_\sigma$, respectively, satisfying
property~\eqref{eq:invariance} 
then indeed 
\[ (R_h \otimes R_h) \cdot (\alpha_\rho \otimes \alpha_\sigma) 
= (R_h \alpha_\rho) \otimes (R_h \alpha_\sigma) 
= \alpha_\rho \otimes \alpha_\sigma \] 
and the resulting mixed state on $V_\rho \otimes V_\sigma$ is invariant under right (diagonal) multiplication by $h$.

Returning to the case where $G = S_k^n$ and 
\[
H = \{ \id, {\bar{m}} = (m_1, \ldots,m_k)\}\enspace,
\]
note that if $\overline{\rho} = \rho_1 \otimes \cdots \otimes \rho_n$ is a
one-dimensional representation of $G^n$ in which the number of
$\rho_i$ isomorphic to the sign representation is odd, then 
${\bar \rho}(\id) = 1$ while ${\bar \rho}(\bar{m})$ = -1.  Thus the projection operator
$\Pi_H^{\bar \rho} = (1/2)(R_1 + R_m)$ is the \emph{zero operator}, and $\bar \rho$ is a 
missing harmonic in the sense of~\cite{Banff}.  Since there are no
mixed states $\alpha$ over $V_{\bar \rho}$ with the property that $R_h \alpha
= \alpha$ for all $h \in H$, such a $\bar \rho$ can never be
observed in the course of the sieve process. This establishes the
statement in ingredient (3).

\subsection{Determining the hidden subgroup}
\label{sec:determining}

We now describe how the above algorithm, which determines only if the hidden
subgroup is the trivial subgroup, can be used to determine $H$ exactly
in the case $H = \{\id, \bar{m}\}$. For each $i$ from $1$ to $n$,
and each odd involution $b \in S_k$, we will apply our algorithm to determine
if $m_i = b$. To accomplish this, we run the algorithm with the modified
oracle $f_i^b$ defined by
$$
f_i^b(\bar g) = (f(\bar g), c_b(g_i))\enspace,
$$
where $c_b(g_i)$ returns the coset representative of the coset of $\{1, b\}$
containing $g_i$. If $m_i = b$, then $f_i^b$ will be distinct and constant
on the cosets of $H$, and the algorithm will output that the subgroup
is nontrivial. Otherwise, $f_i^b$ will be one-to-one and the algorithm
will output that the subgroup is trivial. Repeating this process
for every $i$, and every $b$, we can fully determine $\bar{m}$.

We have assumed throughout this section that when $H$ is nontrivial, every coordinate of the hidden involution $\bar{m}$ is nontrivial. We show in Section~\ref{sec:general-case} how to determine which coordinates are nontrivial and so relax this assumption.

\section{Generalizing to other base groups}\label{sec:generalize}

\subsection{High-dimensional missing harmonics}

While our exposition above focused on symmetric base groups $G = S_k$, 
our algorithm can be generalized to a much larger family of groups. Most generally,
we consider base groups $G$ with an involution $\mu$ and an irrep $\rho$
such that $\rho(\mu) = -\identity$. This turns out to be equivalent to a nice
group-theoretic condition, as the following proposition shows:

\begin{prop} \label{condition-thm} 
  Let $G$ be a group with an involution $\mu$. The following statements are equivalent:
\begin{enumerate}
\item There is a $\rho \in \hat G$ such that $\rho(\mu) = -\identity$.
\item There is a normal subgroup $N$ of $G$ such that $\mu \notin N$ and $\mu N \in Z(G/N)$.
\end{enumerate}
\end{prop}
\begin{proof}
First, assume (1) and let $N$ denote the kernel of $\rho$. Then certainly $\mu \notin N$.
Also, $G/N \cong \text{Im}\ \rho$ via the isomorphism $gN \mapsto \rho(g)$, 
and $\rho(\mu) = - \identity \in Z(\text{Im}\ \rho)$. 

Conversely, assume (2) and choose any irrep $\rho'$ of $G/N$ which takes a nontrivial value
on $\mu N$. We can pull $\rho'$ back to a $G$-irrep $\rho$ via $\rho(g) = \rho'(gN)$.
Since $\mu N \in Z(G/N)$, $\rho(\mu) = \rho'(\mu N) = \lambda \identity$ by Schur's Lemma, 
and since $\mu^2 = 1$, we have $\lambda = -1$.  
\end{proof}

With base group $G$ as above, our general algorithm will identify subgroups of
the form $H = \{ \id, \bar{m} \} \leq G^n$ where each $m_i$ is $1$ or a conjugate of $\mu$.
Our algorithm for $S_k^n$ above relied on the fact that the sign representation of $S_k$ 
is one-dimensional. However, given the condition of Proposition~\ref{condition-thm}, 
the dimension of $\rho$ could certainly be greater than one. This problem is easily
overcome by treating this Simon's problem over $K^n$ rather than $G^n$, 
where $K$ is the \emph{projective kernel} of $\rho$, i.\,e., the subgroup of $G$ consisting of 
elements $g$ such that $\rho(g)$ is a scalar.  Indeed, as $K$ is normal, it contains the 
conjugates of $\mu$, and hence $H \leq K^n$.  Furthermore, the restriction of $\rho$ to $K$ 
decomposes into $d_\rho$ copies of some one-dimensional irrep of $K$, 
and thus
$$
\text{Res}_N^G \rho (g) \cong \bigoplus_{i=1}^{d_\rho} \chi(g)
$$
for some character $\chi$. In particular, $\rho(\mu) = -\identity$ implies $\chi(\mu) = -1$, 
reducing to the previous situation where the missing harmonic is one-dimensional.  

\subsection{One-dimensional missing harmonics and determining nontrivial coordinates}
\label{sec:general-case}

Let $G$ be a group, let $\mu$ be an involution, and let $G'=[G,G]$ denote the commutator subgroup of $G$. Since $G/G'$ is abelian, the existence of a character $\chi$ of $G$ such that $\chi(\mu) = -1$ is equivalent to the condition $\mu \notin G'$.
Under this condition on the base group $G$, we consider the HSP over  $G^n$ where the hidden 
subgroup is $H = \{ \id, \bar{m} \}$, where each $m_i$ is either $1$ or a conjugate of $\mu$. 
Indeed, this is the case in the original Simon's problem, where $\mu$ is the nonidentity element of $\ZZ_2$, 
and $\chi$ is the nontrivial character. 

To adapt our algorithm to the case described above, we make one simple modification.
For the symmetric base group case, we paired $S_k^n$-irreps $\bar \rho$ according to a rule which attempted
to pair $S_k$-factors $\rho_i$ with either $\rho_i^*$ or $\rho_i^\perp \cong \rho_i^* \otimes \pi$, 
where $\pi$ is the sign character of $S_k$. Instead, we will now pair a $G$-factor $\rho_i$
with $\rho_i^* \otimes \psi$, where $\psi$ is a one-dimensional representation of $G$ selected uniformly at random.
Note that since $\rho_i \otimes \rho_i^*$ contains a copy of the trivial representation,
$\rho_i \otimes \rho_i^* \otimes \psi$ will contain a copy of $\psi$, and hence each pairing 
made in this fashion will allow our algorithm to make progress, as before. The
final collection $\Lambda_\text{end}$ of irreps will thus consist entirely of one-dimensional representations
of $G^n$. To see how we can determine the nontrivial coordinates of $\bar{m}$, we now show that 
these final irreps will in fact be uniformly random elements of 
$$
H^\perp = \left\{\psi \in \widehat{G}^n : \psi\;\text{one-dimensional}, H \subset \ker \psi \right\},
$$
i.\,e. the one-dimensional representations of $G^n$ which take the value $1$ on $H$. Since characters
outside $H^\perp$ take the value $-1$ on $\bar{m}$, they cannot appear in $\Lambda_\text{end}$
since they are missing harmonics, as before. 
It thus suffices to show that 
the distribution which dictates the appearance of irreps in $\Lambda_\text{end}$
is invariant under multiplication (tensor product) by any element of $H^\perp$. 
We accomplish this by demonstrating, given $\psi \in H^\perp$ and 
$\bar \rho \in \Lambda_\text{end}$, an ``alternate history'' of the algorithm which instead produces 
$\bar \rho \otimes \psi$ with the same probability. 

We first remark that the formalism of quantum mechanics does not distinguish between the state $\kett{\phi}$ and $\lambda \kett{\phi}$, when $\lambda \in \CC$; indeed, they determine the same density matrix. In particular, letting $A_\chi$ be the operator $\sum_g \chi(g) \kett{g}\bbra{g}$ we have $A_\chi \kett{cH} = \chi(c) \kett{H}$ whenever $\chi \in H^\perp$; thus $\rho_H$ is invariant under $A_\chi$ for any $\chi \in H^\perp$. Note, furthermore, that for any element $f \in \CC G$,
\begin{equation}\label{eq:shift}
\widehat{f}(\rho \otimes \chi) = \sum_{g} \rho(g)\psi(g)f(g) = \widehat{A_\chi f}(\rho)\enspace.
\end{equation}
Evidently, for $\chi \in H^\perp$ the representations $\rho$ and $\rho \otimes \chi$ are equally likely to appear as initial irreps in our sieve and, moreover, the mixed state observed in these two cases is equivalent (under the natural isomorphism of the vector spaces on which $\rho$ and $\rho \otimes \chi$ operate).

The sieve history of our final $\bar \rho \in \Lambda_\text{end}$ is described by 
a (directed) binary tree whose vertices are decorated with the $G^n$-irreps measured at the point when the two children were combined; the leaves are labeled by initially sampled irreps. We also associate with each vertex the mixed state produced by sampling the states of the children. Our alternate history 
is described simply by choosing any (directed) path that begins with $\bar \rho$
and ends with some initially sampled irrep, and applying the tensor product by $\psi$ to the label of each vertex along that path; as $\chi \in H^\perp$, it is one-dimensional, and the labels along the path still correspond to irreps. The rest of the tree remains unchanged.

Recall that each internal node in the tree corresponds to an application of quantum ingredient (2), 
which produces a new child state in an isotypic subspace $W_\tau$ from parents lying in 
irreps $\bar \rho$ and $\bar \sigma$. Observe, first of all, that if $\bar \rho \otimes \bar \sigma = \bigoplus \bar\tau$ and $\chi \in H^\perp$, then  $\bar \rho \otimes (\bar \sigma \otimes \chi) = \bigoplus \bar\tau \otimes \chi$. At the very least, this new labeling of the tree is consistent with decomposition of tensor products into their constituent isotypic spaces. Note, also, that as the matching algorithm pairs a representation $\rho$ uniformly with representations $\rho \otimes \psi$ (for one-dimensional $\psi$), the probability associated with each ``combine'' step is unchanged.

Finally, note that if $\bar{\rho}$ and $\chi$ are representations of $G^n$, with $\chi$ one-dimensional as above, then there is an isomorphism between the space on which $\bar{\rho}$ acts and the space on which $\bar{\rho} \otimes \chi$ acts which preserves the decomposition of these spaces into irreducible subspaces. For concreteness, we may treat these spaces to be the same in the original and alternate trees above (where the only difference in $G$'s action on the subspaces is by a scalar given by $\chi$); in particular, the projection operators associated with the measurements occurring at each combine step are identical. By~\eqref{eq:shift}, the mixed state appearing at the leaf involved in the relabeling is also unaffected by this process, as desired. Hence, the final irreps are selected uniformly at random from $H^\perp$. Since we required that $\mu \notin G'$, we can now efficiently determine the nontrivial coordinates of $\bar{m}$ by linear algebra, just as in Simon's original algorithm.

\subsection{Examples}

The case described in the previous section can provide us with a large family of familiar examples
of base groups. If $G$ has a subgroup $K$ of index $2$, then $K$ is normal with $G/K \cong \ZZ_2$. 
It follows that $G$ has a one-dimensional representation $\pi$, analogous to the sign representation 
in $S_k$, which is $+1$ on $K$ and $-1$ on the nontrivial coset of $K$.  Thus $\pi$ is a missing 
harmonic for any involution $\mu \notin K$, and we can distinguish subgroups of the form 
$H = \{ \id, \bar{m} \}$ for $m = (m_1, m_2, \dots, m_n)$
where each $m_i$ is $1$ or an involution outside $K$. In the case $G = S_k$ studied above, 
$K$ is the alternating subgroup $A_k$ and $\pi$ is simply the sign representation. Similarly, 
if $G$ is a dihedral group $D_k$, the normal subgroup $K \cong \ZZ_k$ 
consists of the rotations.  Finally, if $G$ is a wreath product of the form $L \wr \ZZ_2$, then 
$K \cong L \times L$.  For a detailed description of the irreps of $G$ in this case, see ~\cite{AMR05}.

\end{document}